\documentclass[intlimits,twoside,a4paper]{article}

\usepackage[cp1251]{inputenc}

\usepackage[eqsecnum]{cmpj3}

\RequirePackage{color}

\issue{2024}{27}{3}{33603}
\doinumber{10.5488/CMP.27.33603}
\title[Ising model with varying spin strength on a scale-free network]%
{Ising model with varying spin strength on a scale-free network:  scaling functions and critical amplitude ratios}
\author[M. Krasnytska]{M. Krasnytska\orcid{0000-0002-0464-5741}\refaddr{label1,label2,label3}\thanks{E-mail: \email{kras$\_$mariana@icmp.lviv.ua}.}}
\addresses{
\addr{label1} Institute for Condensed Matter Physics of the National Academy of
Sciences of Ukraine, 79011, Lviv, Ukraine
\addr{label2}  ${\mathbb L}^4$ Collaboration \& Doctoral College for the Statistical Physics of Complex Systems,
Leipzig-Lorraine-Lviv-Coventry, Europe 
\addr{label3} Haiqu, Inc., Shevchenka Str., 120 G, 79039, Lviv, Ukraine
}

\Keywords{phase transitions, Ising model, universality, scaling functions, critical amplitude ratios, complex network}
\date{Received January 21, 2024, in final form March 20, 2024}
\begin{document}

\maketitle

\begin{abstract} 
Recently, a novel model to describe ordering in systems comprising agents which, although
matching in their binarity (i.e., maintaining the iconic Ising features of ``+'' or ``--'', ``up'' or ``down'',
``yes'' or ``no''), still differing in their strength was suggested [Krasnytska et al., J. Phys. Complex., 2020, \textbf{1}, 035008]. The model was analyzed for a particular case
when agents are located on sites of a scale-free network and agent strength is a random
variable governed by a power-law decaying distribution. For the annealed network, the exact
solution shows a rich phase diagram with different types of critical behavior and new universality
classes. This paper continues the above studies and addresses the analysis of scaling functions and universal critical amplitude ratios for the model on a scale-free network.
%
%
%\keywords Up to six keywords (\href{https://physh.aps.org/browse}{Physics Subject Headings})
\printkeywords
\end{abstract}

{\centering{Like after the winter the green spring comes,\\
        the strong bright positive emotions \\
        will emerge when hearing \\
      {\bf ``Ralph Kenna''!\footnote{On the occasion of 60th birthday and memorials of  RALPH KENNA ---  Teacher,  Colleague and  Friend}} \\
       YOU always will be here for us and with us!\\}}
 
\section{Introduction, motivation}\label{I} 

The motivation to explore new spin models follows from enhancing our understanding of complex systems and ordering phenomena, extending beyond conventional boundaries.  Spin models have proven to be invaluable tools in studying a wide range of systems, particularly within the realm of statistical mechanics. However, as our knowledge of complex systems and diverse agent-based systems has grown, there arises a need for novel models that can better capture the intricacies of these scenarios. Recently, we proposed an extension of the Ising model \cite{Krasnytska20,Krasnytska21}: it keeps the binary nature of the Ising model while relaxing the requirement of fixed spin length on each site. In our model, the length of each spin $\mathcal{S}$ is treated as a quenched random variable, following a given probability distribution function $q(\mathcal{S})$:
 \begin{equation}\label{q(S)}
q(\mathcal{S})=c_{\mu}\mathcal{S}^{-\mu}, \hspace{1em} \mathcal{S}_{\rm min} \leqslant
\mathcal{S} \leqslant\mathcal{S}_{\rm max},
\end{equation}
where $c_\mu$ is a normalizing constant.

It was shown that the decay exponent $\mu$ plays a role of a global parameter and determines the collective behavior.  The Model\footnote{Here and below by the definition 'the Model' we  define Ising model with varying spin strength.}  is similar to other spin
models that are used to study the impact of structural disorder on collective behavior  \cite{Mattis1,Mattis2,Hopfield2,Hopfield3,spin_glasses1,spin_glasses2,Folk03}.

\begin{figure}[h!]
\includegraphics[width=0.45\columnwidth]{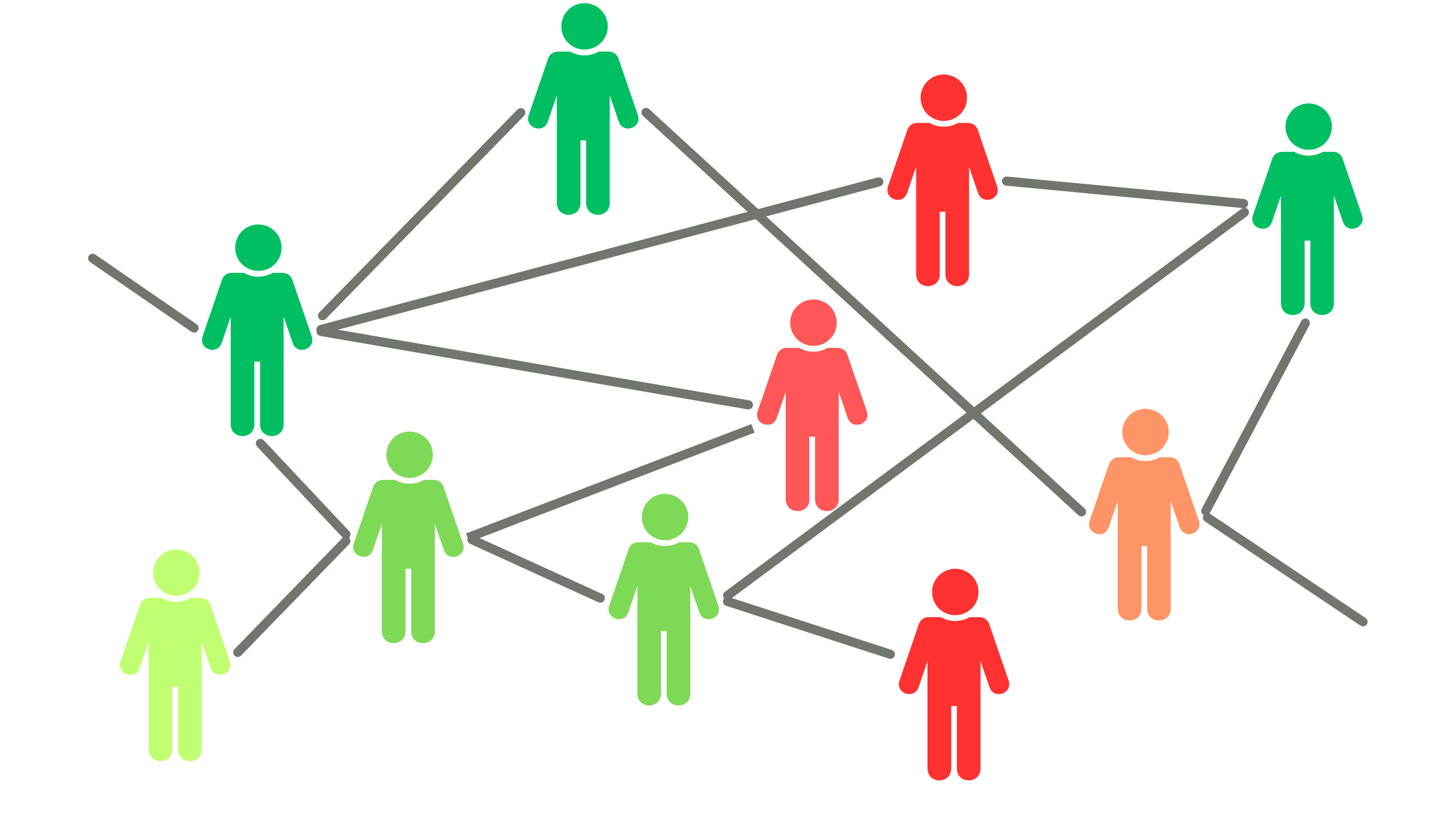}
\hspace{1cm}
\includegraphics[width=0.45\columnwidth]{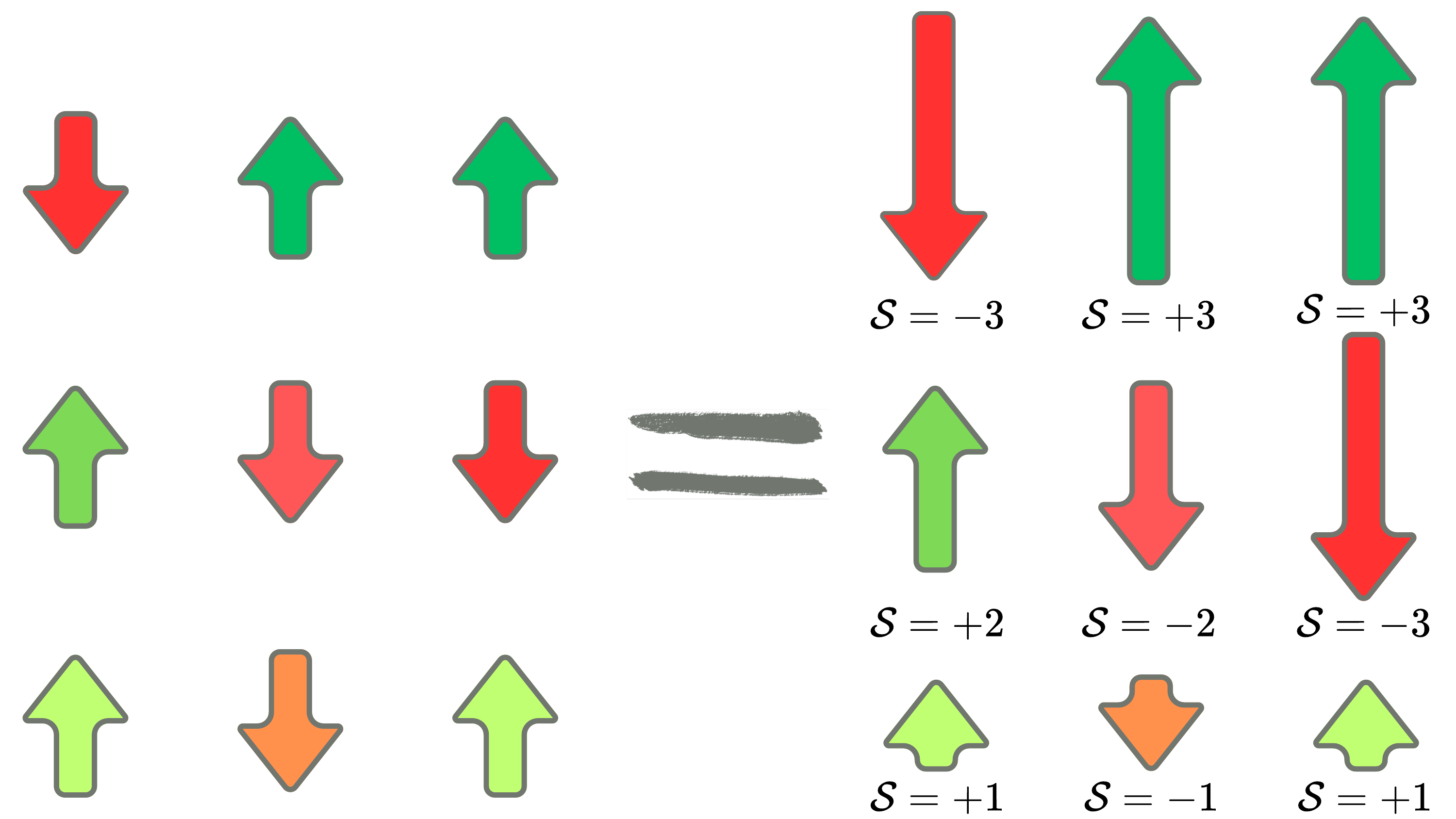}
\centering (a) \hspace{7cm}(b)
\caption{(Colour online) A schematical representation of the  Model. (a) The model on a social network. For each individual of the network, the ``spin'' property is assigned. E.g. considering the network for positive and negative emotions spreading, the nodes of the graph may obey either positive spin values (spin up, $+\mathcal{S}$) --- green icons, or negative (spin down, $-\mathcal{S}$) --- red icons. The intensity of the color corresponds to the spin strength (spin length). (b) The model on a lattice. Similarly, to the social spin model, the magnetic systems contain binary spins (positive (up) --- green,  and negative (down) --- red) which differ in the strength of their expression (color intensity or spin length $\mathcal{S}=1,2,3$). }  
\label{fig1}
\end{figure}

Our primary motivation was to investigate critical behavior 
of the Model on different graphs to analyze the collective behavior within complex networks \cite{Holovatch17,Dorogovtsev08}.
The main focus was on investigating the concurrence of individual (spin)  and global (graph's topology) properties. In figure~\ref{fig1}(a) an example of the Model for a social network is presented. Traditional spin models often fail to explain the dynamics and opinion formation. This model, incorporating individual spin strength captures the intricate nature of social networks but also may have practical applications in marketing, politics, and game theory. {As a special case of the interdisciplinary application of the model, an example can be drawn from neuroscience. Cognition rests on connectivity structures in the brain while neural communication depends on an interplay between its network geometry and topology \cite{Lynn19,Seguin18}. The variability in nodal strengths may contribute to network modelling \cite{Betzel17}, accommodating differences in how the nodes, defined as brain tissue collections~\cite{Stanley13}, operate within the brain networks. Nodes, acting as interactive units, may exhibit strength variations reflecting internal structures \cite{Park13} or changes in functionality over time~\cite{Bagarinao19}.}

On the other hand, for the statistical physics objects, the Model offers insights into the study of the impact of the structural disorder on collective behavior in magnetic systems with polydisperse elementary moments \cite{Tadic1,Tadic2}.  
Using a field-theoretical renormalization group approach to examine the renormalization group flow under various initial conditions we computed effective critical exponents. Our findings illustrate how the variation in properties of the magnetic components governs the effective exponents. It was shown that the model with the presence of two (and more) chemically different magnetic components (spins with different lengths) is equivalent to the diluted magnets with non-magnetic impurities  \cite{Dudka23}. In addition, we intend to validate the observed behavior through Monte Carlo simulations.

In our previous studies of the Model, we have explored the free energy and phase diagram in different regions of the global parameters. We obtained the asymptotics for the thermodynamic characteristics of a system that appear near the critical point. Critical exponents and logarithmic correction exponents, which control the behavior of different observables close to the critical point, are among these characteristics. However, there are other universal characteristics to quantify the critical behavior:  if systems belong to the same universality class, they not only have the same values for critical exponents but also share identical critical amplitude ratios and scaling functions \cite{Privman91}.
The motivation to study the scaling functions has both theoretical \cite{Stanley72,Stanley99,Delfino98}   
and practical applications ---
critical amplitudes and scaling functions are reachable both by Monte Carlo simulations  \cite{CASELLE1998613,ENGELS2003277,Kenna11} and experimental measurements \cite{Davies83,STOOP91,McLachlan98}.
Of high importance is the analysis of scaling functions and amplitudes for systems with a large number of interacting agents to predict their behavior and critical properties, facilitating the design and optimization of magnetic and macromolecular materials for various applications \cite{Dulan19,Krummenacher22}. Besides, the motivation to study those characteristics on different types of graphs for a better understanding of the processes in social systems and magnetic materials (mentioned above) is valid for macromolecular systems which are inherently complex, with numerous interacting components. Scaling functions offer a powerful tool for describing the collective behavior of these components during phase transitions. Analyzing the scaling functions allows us to describe the dynamics of macromolecular transformations.

The plan of the rest of the paper is as follows. In section~\ref{II} we describe the Model under investigation, detailing its critical bahaviour and the phase diagram. We list the main equation for the free energy to be analysed.  In the next section~\ref{III} the definitions of amplitude ratios and scaling functions are presented. Section~\ref{IV} presents the outcomes and application of the scaling theory to the Model in three different regions of the second-order phase transition with different universality classes. The concluding section synthesizes the key findings, draws connections to the research objectives, and discusses possible implications of the results.

 \section{The model: free energy and phase diagram}\label{II}

\begin{figure}[t!] 
\centering
\includegraphics[width=0.85\columnwidth]{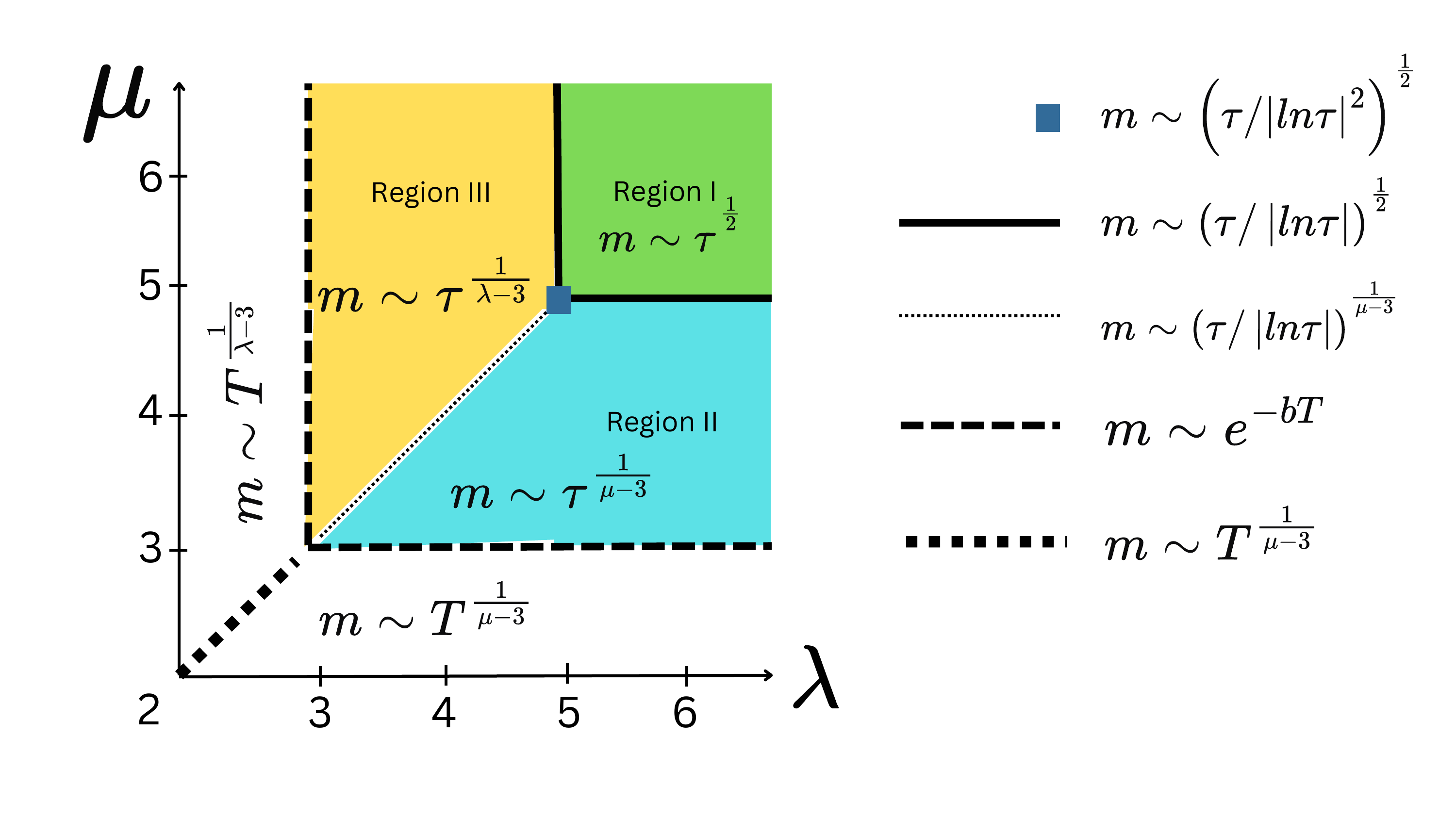} 
\caption{(Colour online) Phase diagram of the Model on a scale-free network in  $\mu$-$\lambda$ plane.  The asymptotics for magnetization $m$ behaviour in different regions $\mu$ and $\lambda$ are shown. The logarithmic corrections and the asymptotics in ordered states are shown by different line types. Three different sets of logarithmic corrections to scaling appear for the model. Three colored regions on the phase diagram correspond to three different universality classes of critical exponents (see \cite{Krasnytska20,Krasnytska21} for more details).}  
\label{phaseDiagram}
\end{figure}

The Hamiltonian of the  
model reads \cite{Krasnytska20,Krasnytska21}:
\begin{equation}\label{9}
{\cal H} = -\frac{1}{2}\sum_{i\neq j} J_{ij} S_i S_j
- h \sum_i S_i \, , \hspace{1cm} S_i=\pm
\mathcal{S}_i\, ,
\end{equation}
with the spin strength power-law distribution (\ref{q(S)}). Here, $h$ is an external magnetic field and $J_{ij}$ --- an adjacency matrix with matrix elements $0$ (if there are no links between nodes $i$ and $j$) and $1$ (nodes are connected). The probability $p_{ij}$ of two nodes being connected is determined by graph properties. E.g., for a complete graph all nodes are connected with each other, so $p_{ij}=1$, for Erd\H{o}s--R\'eny graphs the probability of links is fixed to some random value $p_{ij}=c$ ($0<c<1$). And the last but not least case should be mentioned and the one we consider here -- scale-free network,  where for an annealed network $p_{ij}=k_ik_j/(N <k>)$.  
Each
individual is represented as a complex network node of a given degree $k_i$ (i.e., a number of persons
connected to it via social links) and of a given strength $S_i$.  
The node degree  distribution function for a scale-free network is determined by a power-law decay:
\begin{equation}\label{p(k)}
p(k)=c_{\lambda}k^{-\lambda}.
\end{equation}

For the Model {\bf{on a scale-free network}}, the free energy was obtained in the form \cite{Krasnytska20,Krasnytska21}:
\begin{equation}\label{IV.2}
\Phi_{\mu,\lambda}(m,T,h) = \frac{\langle k \rangle m^2T}{2}-c_\mu
c_\lambda
m^{\frac{\lambda+\mu-2}{2}}I_{\lambda,\mu}(\sqrt{m})
-\frac{\langle \mathcal{S}^2\rangle \langle k\rangle}{T}mh \, .
\end{equation}
where
\begin{equation}\label{III.6a}
I_{\lambda,\mu}(m)=
\int_{\sqrt{m}}^{\infty}\int_{\sqrt{m}}^{\infty} \frac{\ln
\cosh( k\mathcal{S})}{k^\lambda \mathcal{S}^\mu} \, \rd\mathcal{S}\rd k,
\end{equation} 
and $m$ is an order parameter, $\langle k \rangle$ is the average node degree, $\langle \mathcal{S}^2\rangle$ is the average of the squared spin strength.   

The resulting free energy is symmetric upon an interchange of
indices $\mu\leftrightarrow \lambda$. The leading asymptotics for the free energy are determined by a smaller parameter between $\mu, \lambda$ (the 'fatter' tail of the distribution function wins the competition in defining the universality class).
So we can consider 3 main cases:   $\mu>\lambda$,   $\mu=\lambda$,  $\mu< \lambda$. Peculiar phenomena emerge in the regions with $\mu=\lambda$, where the changes in critical exponent $\mu$- or $\lambda$- dependencies occur. Two new universality classes for logarithmic correction exponents are observed here. The critical behaviour of the model is summarized in figure~\ref{phaseDiagram}. There are two global parameters --- $\mu$ and $\lambda$ --- that affect the critical behaviour. However, for further analysis, namely scaling functions and amplitude ratios, we skip the regions with ordered phase\footnote{{For the Model, when $2<\lambda\leqslant3$ or $2<\mu\leqslant3$, the system is ordered at any finite temperature and the magnetization behaves either exponentially or as a power law of the temperature $T$ (see asymptotics in figure~\ref{phaseDiagram})}} and second order phase transition lines with logarithmic corrections (see asymptotics in figure~\ref{phaseDiagram}). We  focus on the free energy expressions in three main regions: Region I (MFA), Region II ($\mu$-dependent critical behaviour) and  Region III ($\lambda$-dependent critical behaviour): 
\begin{align} \label{region1}
\Phi (m,\tau,h) &=  \frac{A}{2}\tau m^2 + \frac{B}{4} m^4-Dmh, & &\text{(Region I)} ,\\
%\end{equation}
%\begin{equation}
\label{region2}
 \Phi (m,\tau,h) &=  \frac{A}{2}\tau m^2 + \frac{B'}{4} m^{\mu-1}-Dmh, & & \text{(Region II)} , \\
%\end{equation}
%\begin{equation}
\label{region3}
 \Phi (m,\tau,h) &=  \frac{A}{2}\tau m^2 + \frac{B''}{4} m^{\lambda-1}-Dmh,
&&\text{(Region III)} ,
 \end{align}
where $\tau=|T-T_c|/T_c$ and coefficients are all positive (their exact expressions can be found in appendices~\cite{Krasnytska21}).

 \section{Definitions: amplitude ratios and scaling functions}\label{III}

We 
observe the power-law asymptotic behaviour for thermodynamic characteristics in the region of the second order phase transition near the critical point $T=T_\mathrm{c},\,
h=0$\cite{Stanley99,Privman91}. 
For the  order parameter $m$, isothermal susceptibility
$\chi_T$, specific heat $c_h$ and magnetocaloric coefficient $m_T$,
 the following   functions of
$\tau={|T-T_\mathrm{c}|}/{T_\mathrm{c}}$ can be written in terms of critical amplitudes and critical exponents:
\begin{equation}\label{amplitude1}
m=B_-\tau^\beta,\qquad \chi_T =\Gamma_\pm \tau^{-\gamma}, \qquad
c_h=\frac{A_\pm}{\alpha}\tau^{-\alpha},\qquad m_T=B_T^{\pm} \tau^{-\omega}
\qquad \text{at} \qquad h=0.
\end{equation}
 Exactly at critical point $T=T_\mathrm{c}$ (i.e.,
$\tau=0$) the next asymptotics are valid:
\begin{equation}\label{amplitude2}
m=D_\mathrm{c}^{{-1}/{\delta}}h^{1/\delta}, \qquad \chi=\Gamma_\mathrm{c}
h^{-\gamma_\mathrm{c}}, \qquad c_h=\frac{A_\mathrm{c}}{\alpha_\mathrm{c}}h^{-\alpha_\mathrm{c}},
\qquad m_T=B^c_T h^{-\omega_\mathrm{c}} \qquad \text{at}
\qquad \tau=0.
\end{equation} 

Unlike the critical exponents, the critical amplitudes are
non-universal, but their ratios (combinations) appear to be universal \cite{Privman91} and are used for the analysis
below:
\begin{equation}\label{amplituderatios}
R_\chi^\pm=\Gamma_\pm D_\mathrm{c} B_-^{\delta-1}\,, \quad R_\mathrm{c}^\pm=\frac{A_\pm \Gamma_\pm}{\alpha B_-^2}\,,\quad
R_A=\frac{A_\mathrm{c}}{\alpha_\mathrm{c}} D_\mathrm{c}^{-(1+\alpha_\mathrm{c})}B_-^{-2/\beta}\,,
\quad A_+/A_-\,, \quad \Gamma_+/\Gamma_-\,.
\end{equation}

To start with the analysis of scaling functions we should refer to the scaling hypothesis for the Helmholtz free energy
$F(\tau,m)$, which states that this thermodynamic potential is a generalized
homogeneous function \cite{Stanley72}. So using the scaling variable $x$, the function of two variables $F(\tau, m)$ can be written as some function of $f(x)$: 
\begin{equation}\label{scalingf}
F(\tau,m)\approx \tau^{2-\alpha} f_{\pm}(x),
\end{equation}
where 
\begin{equation}\label{x}
x=m/\tau^\beta 
\end{equation}
is a scaling variable, $f_\pm(x)$ is a scaling function, signs
$+$ and $-$ correspond to $T>T_\mathrm{c}$ and $T<T_\mathrm{c}$, respectively.  All other thermodynamic potentials are also generalized homogeneous functions.
Based on the expression for the free energy one can also represent
the thermodynamic functions in terms of appropriate scaling functions. The scaling functions for the magnetic field, entropy, heat capacity, isothermal
susceptibility, and magnetocaloric coefficient are defined as follows (see
e.g.,~\cite{Stanley72,vonFerber11}):
\begin{eqnarray}  \label{97}
 h(m,\tau) &=& \tau^{\beta \delta}H_\pm (x), \\ \label{99}
S(m,\tau) &=& \tau^{1-\alpha} {\cal S}(x), \\
\label{100} 
 c_h(m,\tau) &=&(1\pm \tau)\tau^{-\alpha} {\cal C_\pm}(x),
  \\ \label{101}
 \chi_T(m,\tau) &=&\tau^{-\gamma}\chi_\pm(x),  \\ \label{102}
 m_T(m,\tau) &=& (1\pm\tau)\tau^{\beta-\gamma}{\cal M}_\pm(x).
\end{eqnarray}

 \section{Results}\label{IV}

In this section, we complete the previous description of the critical behaviour of the Model on a scale-free network by calculating its amplitude ratios and scaling functions
in the vicinity of critical point at different second order phase transition regimes.

\subsection{Critical amplitude ratios}\label{subsec4.1}

\begin{figure}[!t]
\centering
\includegraphics[width=0.8\columnwidth]{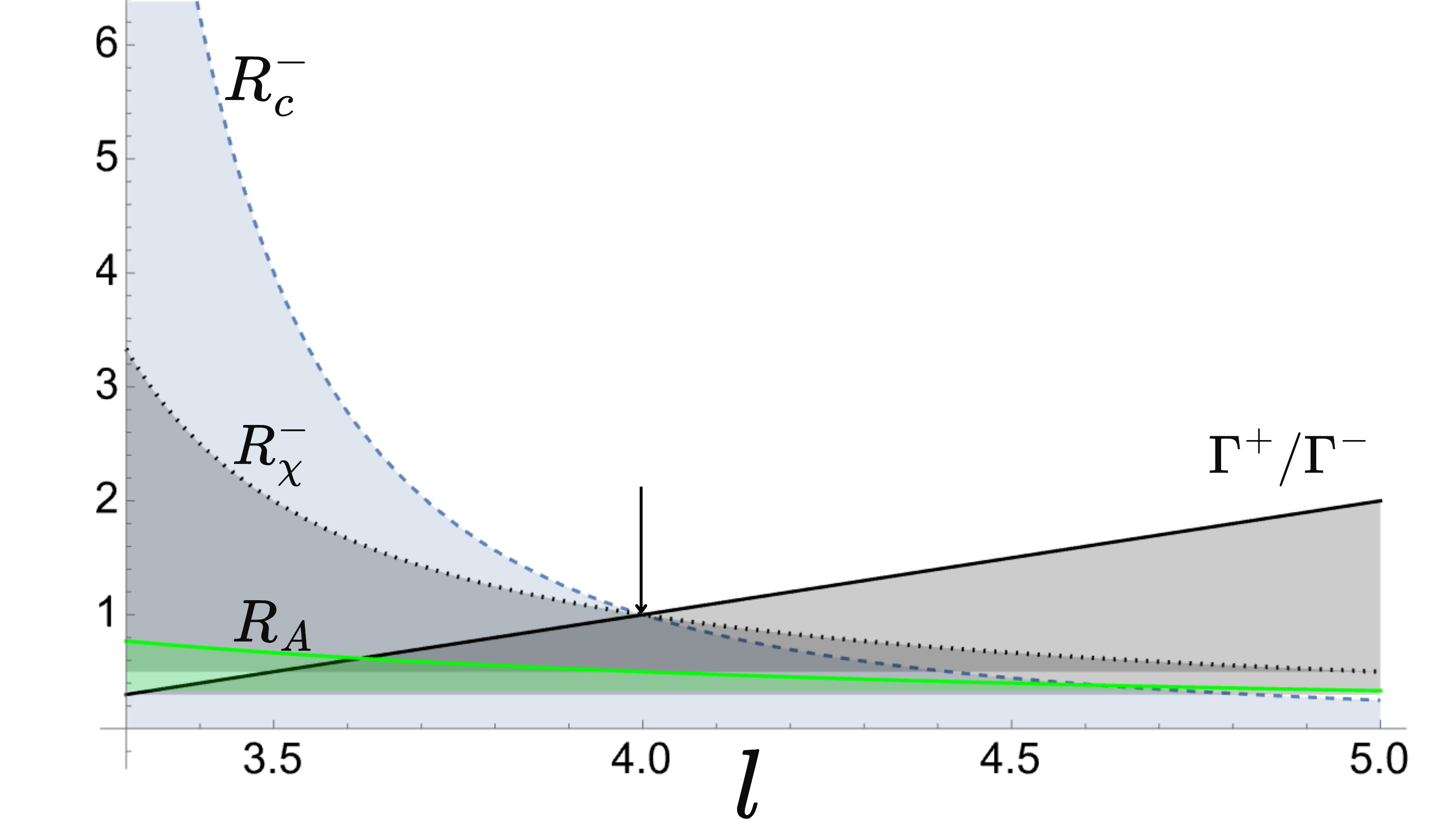} 
    \caption{(Colour online) Critical amplitude ratios for a parameter range $3<l<5$ (here $l$ is either $\mu$ (Region II) or $\lambda$ (Region III)). Solid black line $\Gamma^+/\Gamma^-=l-3$; green line $R_A=1/(l-2)$; dotted $R^-_\chi=1/(l-3)$, dashed $R^-_c=1/(l-3)^2$. For $l\geqslant 5$ the values of amplitude ratios reach MFA values from Region I and remain unchanged. At $l=4$ three functions for amplitude ratios intersect at the same point.} 
     \label{figure1point}
\end{figure}

For analysis, we start from the free energy expressions (\ref{region1})--(\ref{region3}) in three regions of the second-order phase transition. We took the corresponding derivatives over temperature and magnetic field. Thus, thermodynamic characteristics can be obtained and presented in the form of power low functions of $\tau$ and $h$ and critical amplitudes, see (\ref{amplitude1})--(\ref{amplitude2}). Some amplitude ratios remain unchanged in all critical regions (see first three lines for amplitude ratios section in table~\ref{tablescaling}) while the rest depend on the global parameter values and differ in three regions. For Region I, the mean field asymptotic values are obtained. In Region II and III the ratios become $\mu$ and $\lambda$-dependent and in the limit $\mu$, $\lambda\to 5$ approach their mean-field values and remain unchanged for $\mu$, $\lambda \geqslant 5$. There should be mentioned an interesting fact: at $\mu=4$ or $\lambda=4$ three functions for amplitude ratios intersect at the same point, see figure~\ref{figure1point}. It should be mentioned that below $\lambda=4$, we have the last purely imaginary partition function Lee-Yang zero for the Ising model on the scale-free network, which belongs to the same universality class as Region III for the Model \cite{Krasnytska16}. So it makes sense to also analyze the Model in terms of the partition function zeros in order to supplement the critical behaviour. Recently, a preprint on the definition of Lee-Yang and Langer edge singularities from the analytic continuation of scaling functions was published \cite{Karsch23}. 
 
%Do they feel log corrections of Potts model universality class?....... what about at $\lambda=5$? why no special behavior?? it is just a sequence of amplitude ratio definitions... 

\begin{table}[h!]
\caption{Scaling functions and critical
amplitude ratios for the Model on an uncorrelated scale-free network derived from the free energy expression (\ref{region1})--(\ref{region3}) in three regions (see figure~\ref{phaseDiagram}).  Region I: $\lambda,\mu>5$;  Region II: $3<\mu<5$, $\mu<\lambda$;  Region III: $3<\lambda<5$, $\mu>\lambda$. The amplitude ratios, scaling functions and scaling parameter definitions are given by (\ref{amplituderatios}), (\ref{97})--(\ref{102}) and (\ref{x}), correspondingly. \label{tablescaling} }
\vspace{2ex}
\begin{center}
\begin{tabular}{|c|c|c|c|}
\hline
              & Region I  & Region II &  Region III \\ 
              \hline\hline
Amplitude ratios      &   &  &   \\ \hline\hline
$A^+/A^-$   & $0$ & $0$ & $0$\\
$R^+_\chi$   & $1$ & $1$ & $1$\\
$R^+_\mathrm{c}$   & $0$ & $0$ &$0$\\
$\Gamma^+/\Gamma^-$   & $2$ & $\mu-3$ & $\lambda-3$\\
$R^-_\chi$   & $\frac{1}{2}$ & $\frac{1}{\mu-3}$ &$\frac{1}{\lambda-3}$\\
$R^-_\mathrm{c}$   & $\frac{1}{4}$ & $\frac{1}{(\mu-3)^2}$ &$\frac{1}{(\lambda-3)^2}$\\
$R_A$   & $\frac{1}{3}$ & $\frac{1}{\mu-2}$ &$\frac{1}{\lambda-2}$\\[1ex]
              \hline\hline
      Scaling functions       &   &  &   \\ \hline\hline
$f_\pm(x)$   & $\pm \frac{x^2}{2} + \frac{x^4}{4}$ &$\pm \frac{x^2}{2} + \frac{x^{\mu-1}}{4}$& $\pm \frac{x^2}{2} + \frac{x^{\lambda-1}}{4}$\\
$H_\pm(x)$   &  $x^3\, \pm \,x$ & $\frac{\mu-1}{4} x^{\mu-2}\, \pm \, x$ &$\frac{\lambda-1}{4} x^{\lambda-2}\, \pm \, x$\\
${\cal S}(x)$   & $-x^2/2$ & $-x^2/2$ &$-x^2/2$\\
${\cal C_\pm}(x)$   &  $\frac{x^2}{3x^2 \pm 1}$ &$\frac{x^2}{x^{\mu-3}{(\mu-1)(\mu-2)}/{4}\pm 1}$&$\frac{ x^2}{x^{\lambda-3} {(\lambda-1)(\lambda-2)}/{4}\pm 1}$ \\
$\chi_\pm(x)$   &  $\frac{1}{3x^2 \pm 1}$ &$\frac{1}{x^{\mu-3}{(\mu-1)(\mu-2)}/{4}\pm 1}$&$\frac{1}{x^{\lambda-3} {(\lambda-1)(\lambda-2)}/{4}\pm 1}$ \\
${\cal M}_\pm(x)$  & $\frac{x}{3x^2 \pm 1}$ &$\frac{x}{x^{\mu-3}{(\mu-1)(\mu-2)}/{4}\pm 1}$&$\frac{x}{x^{\lambda-3} {(\lambda-1)(\lambda-2)}/{4}\pm 1}$ \\ 
\hline
\end{tabular}
\end{center}
\end{table}

\subsection{Scaling functions}
Passing to the dimensionless energy $F(m, \tau)$ and dimensionless
magnetization $m$  we can   present (\ref{region1})--(\ref{region3}) in three different regions
of the phase diagram (figure~\ref{fig1}) in the following form:
\begin{align}
 \label{f1}
F(m,\tau) &=  \pm \frac{\tau}{2}m^2 + \frac{1}{4} m^4, & &\text{(Region I)} ,\\
%\end{equation}
%\begin{equation}
\label{f2}
 F(m,\tau) &=  \pm \frac{\tau}{2}m^2 +
\frac{1}{4} m^{\mu-1}, & &\text{(Region II)} , \\
%\end{equation}
%\begin{equation}
\label{f2}
F(m,\tau) &= \pm \frac{\tau}{2} m^2 + \frac{1}{4}m^{\lambda-1},
&&\text{(Region III)} ,
 \end{align}
the signs $\pm$ here and in what follows refer to the temperatures
above and below the critical point $T_\mathrm{c}$.

With the expressions for the free energy at hand, it
is straightforward to write down the equation of state and to derive
the thermodynamic functions. The magnetic and entropic equations of
state in the dimensionless variables $m$ and $\tau$ read:
\begin{equation}\label{eqs}
h(m,\tau)=\left.\frac{\partial F(m,\tau)}{\partial m} \right|_\tau \, , \qquad s(m,\tau)=\left.\mp \frac{\partial F(m,\tau)}{\partial \tau}
\right|_m \, .
\end{equation}

The entropic equation of state is obtained by a temperature
derivative at a constant magnetization $m$ while the explicit
$\tau$-dependency is the same in all regions. Therefore, the equation keeps the same form
in all regions on $\mu$--$\lambda$ plane:
\begin{equation}\label{eq8820}
s=-m^2/2, \qquad\qquad \text{(Regions I--III)} .
\end{equation}
Rewriting the free energies (\ref{f1})--(\ref{f2}) in terms of scaling variable $x$ (\ref{x}) the energies can be written in scaling form like (\ref{scalingf}). All other thermodynamic characteristics as a corresponding derivative from the energy can be easily written in terms of the scaling variable. The comprehensive expressions for scaling functions in Regions I, II and III are listed in table \ref{tablescaling}. 
Written explicitly in different regions of $\mu$ and  $\lambda$ the magnetic equation of state and thermodynamic functions $\chi_T$, $c_h$, and $m_T$ differ in Regions I--III. In Region I we obtain the MFA-like behaviour, in Region II and  III the expressions depend on $\mu$ and  $\lambda$, correspondingly. It is easy to observe that in the limit $\mu \to \infty$ ($\lambda \to \infty$) the scaling functions approach theory mean-field expressions from Region I. As mentioned above $S(x)$ does not depend on the global parameters change.

 \section{ Conclusions}\label{V}

The motivation to study new spin models arises from the evolving landscape of scientific inquiry, necessitating adaptable frameworks to comprehend the complexities of diverse systems. These models not only deepen our theoretical understanding but also hold the promise of shedding light on real-world phenomena (social systems and collective behaviour, magnetic materials, etc.)  in a more nuanced and realistic manner. Utilizing critical exponents and scaling functions helps in designing materials with specific properties. This knowledge allows researchers to engineer materials with desirable structural characteristics, providing applications in the fields such as drug delivery, biomaterials, and nanotechnology. The more we get to know about the critical behavior of the system the more effective optimization of the experimental conditions is reached. Applying  scaling laws to social, magnetic, or macromolecular systems aids in optimizing experimental conditions for studies involving phase transitions.  

\begin{table}[h!]
	\caption{The comparative analysis of the critical behavior in different regions of the 2nd order phase transition regime for different Ising models (see the phase diagram in figure~\ref{phaseDiagram}. Model 1. Standard Ising model on a complete graph (Kac or Curie-Weiss model).  Model 2: Standard Ising model on a SF network.  Model 3: Ising model with power-law spin length on a complete graph (or Erd\H{o}s--R\'eny graph); Model 4: Ising model with power-law spin length distribution on a SF network. }
	\vspace{2ex}
    \begin{center}
    	\small
        {%\large
            \begin{tabular}{l|c|c|c|c}
                \hline   & Model 1 \cite{Kac52} & Model 2 \cite{Leone02,Igloi02,vonFerber11}& Model 3\cite{Krasnytska20,Krasnytska21} & Model 4 \cite{Krasnytska20,Krasnytska21}\\
                \hline\hline  
                
                Spin  &    & & $S_i=\pm \mathcal{S}$,   & $S_i=\pm \mathcal{S}$,  \\
                length & $S_i=\pm1$    & $S_i=\pm1$  & $q(\mathcal{S})\sim\mathcal{S}^{-\mu}$  & $q(\mathcal{S})\sim\mathcal{S}^{-\mu}$    \\
                \hline \hline
                Graph &  Complete  & SF network: &  Complete or & SF network: \\
                
                 topology &  graph &  $p(k)\sim k^{-\lambda}$  &  Erd\H{o}s--R\'eny  graph &  $p(k)\sim k^{-\lambda}$ \\
                \hline \hline
                Global &    & &   &  \\
                 parameters  & - & $\lambda$   & $\mu$ & $\lambda$   and $\mu$   \\
                \hline \hline
                 2nd order PT   &  &    &    &    \\
                  regions   & Region I (MFA) & Region I and III  & Region I and II  &  Region I, II and III\\  
                  \hline
                \hline 
            \end{tabular}
        }
    \label{tablemodels}
    \end{center} 
    \end{table}

The critical behavior of the Model considered here is of  special interest because of its wide range of possible applications as well as its universality. The Model is a generalization of the binary spin model (Ising model) and in different regions of parameters reproduces the well-known cases.  Here, we obtained analytical expressions for scaling functions and critical amplitude ratios of the Model. Together, critical exponents and critical amplitude ratios constitute quantitative characteristics of a given universality class and can be reachable in experiments and MC simulations. The results of the paper confirm and fulfill the previous model studies \cite{Krasnytska20,Krasnytska21}. Both global parameters (global network topology parameter $\lambda$ and parameter that defines individual properties of spin strength $\mu$) of the proposed model determine the universality classes. 
{The critical behaviour is determined by the smaller of the two parameters. However, the symmetry of the parameter dependence will change if we select different probability  functions  (e.g., Gaussian) instead of  power laws (\ref{q(S)}) and (\ref{p(k)}). }
 
Last but not least, to demonstrate the universality, problems coverage of the Model and its universality, we present the comparative analysis of a Standard Ising model and Ising model with varying spin strength~\cite{Krasnytska20,Krasnytska21} on different topologies listed in table~\ref{tablemodels}: Model 1. Standard Ising model on a complete graph (Kac or Curie-Weiss model);  Model 2: Standard Ising model on SF network;  Model 3: Ising model with power-law spin length on a complete graph (or Erd\H{o}s--R\'eny graph); Model 4: Ising model with power-law spin length on SF network. The spin values, graph topology, the global parameters are defined in the table for each case.  Also from the phase diagram/free energy analysis, we defined the typical critical behaviour (similar to the regions of the phase diagram in figure~\ref{phaseDiagram}). For Model 1, there is only one 2nd order phase transition regime-MFA behaviour, while for Model 2 and Model 3 there are two critical regions with different scaling behaviour, and for Model 4 we obtain three regions. For Model 3 (Ising model with power-law spin length on complete or Erd\H{o}s--R\'eny graphs),  the free energy is equivalent to that of the standard Ising model ($\sigma_i=\pm1$) on an
annealed scale-free network (Model 2), and the decay exponent~$\lambda$ plays a role of $\mu$ in the current one. So we are able to widen the critical behaviour for the Ising model by introducing a varying spin strength! This was confirmed by critical exponents analysis previously and here by scaling functions and amplitude ratios investigation.

\section*{Acknowledgements}

The work would not be
possible without the tireless and noble work of the Armed Forces of Ukraine.
Also, the work has been done under the support of the National Academy of Sciences of Ukraine, 
%grant for research laboratories/groups of young scientists No 07/01-2022(4),
grant for
research laboratories/groups of young scientists No 07/01-2022(4) 
and National Research Foundation of Ukraine Project 246/0099 ``Criticality of complex systems: fundamental aspects and applications''.
%
%Project 246/0099 ``Criticality of complex systems: fundamental aspects and %applications'' of the National Research Foundation of Ukraine.
%
  My deepest thanks to Ralph Kenna, Yurij Holovatch and Bertrand Berche for all previous discussions and works on the Model, for their comments, corrections and recommendations to the draft of this manuscript.   

\section*{Afterwords}

{\em  Can you imagine who could be Ralph in the Model with varying spin strength?..}

{\em  My response: the person with the \underline{largest positive emotions}, with \underline{lots of connections}, who influences and charges the rest of his \underline{positive Irish energy} and \underline{a great scientist curiosity}!}

\includegraphics[width=0.75\columnwidth]{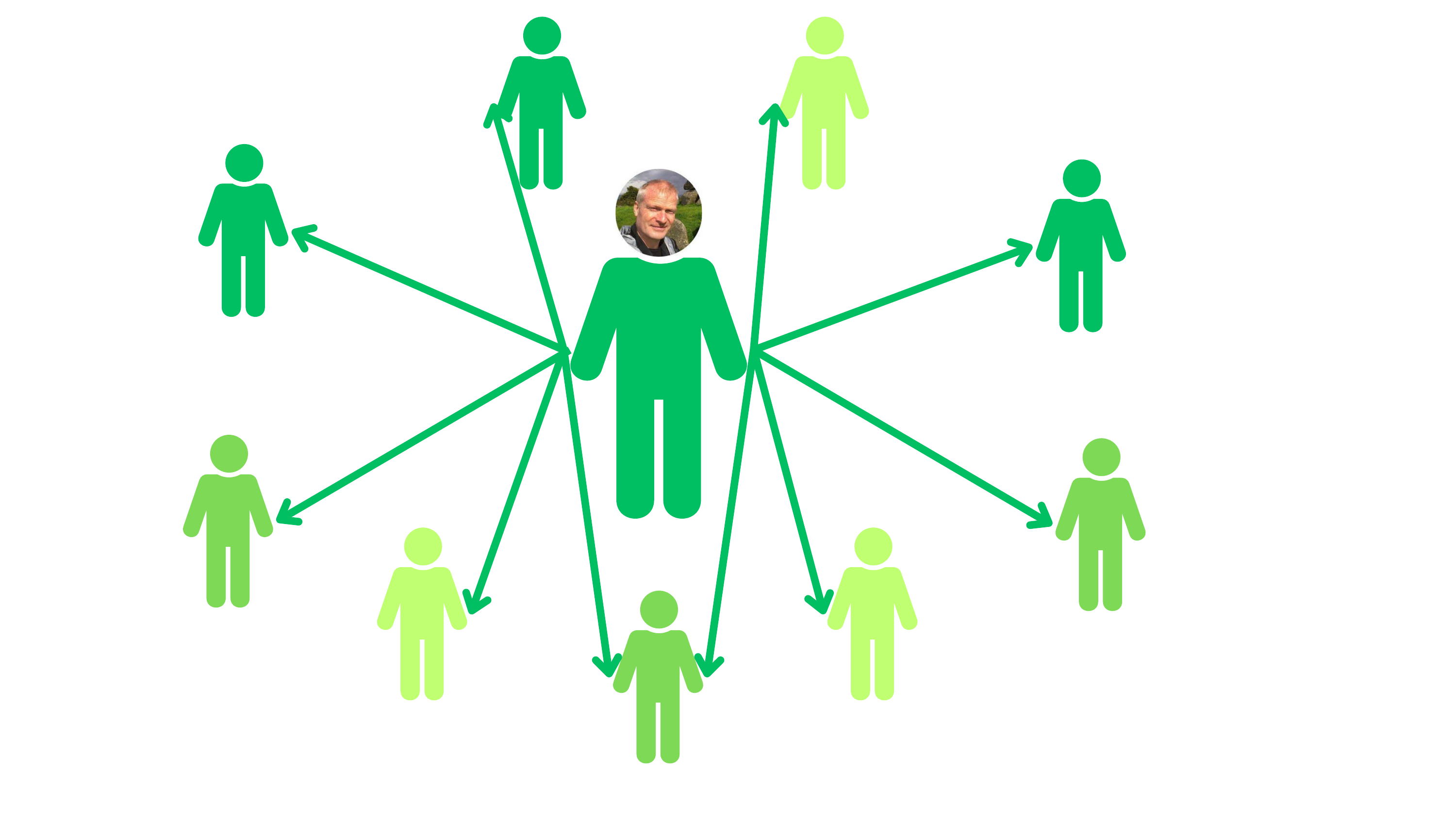}

\bibliographystyle{cmpj}
\bibliography{cmpjxampl}

\begin{thebibliography}{10}
\providecommand{\url}[1]{\texttt{#1}}
\providecommand{\urlprefix}{URL }
\expandafter\ifx\csname urlstyle\endcsname\relax
  \providecommand{\doi}[1]{doi:\discretionary{}{}{}#1}\else
  \providecommand{\doi}{doi:\discretionary{}{}{}\begingroup
  \urlstyle{rm}\Url}\fi
\providecommand{\eprint}[2][]{\url{#2}}

\bibitem{Krasnytska20}
Krasnytska~M., Berche~B., Holovatch~{\relax Yu}., Kenna~R., J. Phys. Complex.,
  2020, \textbf{1}, No.~3, 035008, \doi{10.1088/2632-072X/abb654}.

\bibitem{Krasnytska21}
Krasnytska~M., Berche~B., Holovatch~{\relax Yu}., Kenna~R., Entropy, 2021,
  \textbf{23}, No.~9, \doi{10.3390/e23091175}.

\bibitem{Mattis1}
Mattis~D., Phys. Lett. A, 1976, \textbf{56}, No.~5, 421--422,
  \doi{10.1016/0375-9601(76)90396-0}.

\bibitem{Mattis2}
Bianconi~G., Phys. Lett. A, 2002, \textbf{303}, No.~2, 166--168,
  \doi{10.1016/S0375-9601(02)01232-X}.

\bibitem{Hopfield2}
Pastur~L., Figotin~A., Theor. Math. Phys., 1978, \textbf{35}, 403--414,
  \doi{10.1007/BF01039111}.

\bibitem{Hopfield3}
Hopfield~J., PNAS, 1982, \textbf{79}, 2554--2558, \doi{10.1073/pnas.79.8.2554}.

\bibitem{spin_glasses1}
Mezard~M., Parisi~G., Virasoro~M., Spin Glass Theory and Beyond. An
  Introduction to the Replica Method and Its Applications, Vol.~9, World
  Scientific, Singapore, 1986, \doi{10.1142/0271}.

\bibitem{spin_glasses2}
Dotsenko~V., An Introduction to the Theory of Spin Glasses and Neural Networks,
  World Scientific, Singapore, 1994, \doi{10.1142/2460}.

\bibitem{Folk03}
Folk~R., Holovatch~{\relax Yu}., Yavors'kii~T., Physics-Uspiekhi, 2003,
  \textbf{46}, 169--191, \doi{10.1070/PU2003v046n02ABEH001077}.

\bibitem{Holovatch17}
Holovatch~{\relax Yu}., Kenna~R., Thurner~S., Eur. J. Phys., 2017, \textbf{38},
  No.~2, 023002, \doi{10.1088/1361-6404/aa5a87}.

\bibitem{Dorogovtsev08}
Dorogovtsev~S., Goltsev~A.~V., Mendes~J., Rev. Mod. Phys., 2008, \textbf{80},
  1275--1335, \doi{10.1103/RevModPhys.80.1275}.

\bibitem{Lynn19}
Lynn~C., Bassett~D., Nat. Rev. Phys., 2019, \textbf{1}, 318--332,
  \doi{10.1038/s42254-019-0040-8}.

\bibitem{Seguin18}
Seguin~C., van~den Heuvel~M., Zalesky~A., PNAS, 2018, \textbf{115}, No.~24,
  6297--6302, \doi{10.1073/pnas.1801351115}.

\bibitem{Betzel17}
Betzel~R.~F., Bassett~D.~S., J. R. Soc. Interface, 2017,
  \doi{10.1098/rsif.2017.0623}.

\bibitem{Stanley13}
Stanley~M., Moussa~M., Paolini~B., Lyday~R., Burdette~J., Laurienti~P., Front.
  Comput. Neurosci., 2013, \textbf{7}, \doi{10.3389/fncom.2013.00169}.

\bibitem{Park13}
Park~H.~J., Friston~K., Science, 2013, \textbf{342}, No. 6158, 1238411,
  \doi{10.1126/science.1238411}.

\bibitem{Bagarinao19}
Bagarinao~E., Watanabe~H., Maesawa~S., Sci. Rep., 2019, \textbf{9}, 11352,
  \doi{10.1038/s41598-019-47922-x}.

\bibitem{Tadic1}
Tadi\ifmmode~\acute{c}\else \'{c}\fi{}~B., Malarz~K., Ku\l{}akowski~K., Phys.
  Rev. Lett., 2005, \textbf{94}, 137204, \doi{10.1103/PhysRevLett.94.137204}.

\bibitem{Tadic2}
Tadi{\'{c}}~B., Gupte~N., {EPL}, 2020, \textbf{132}, No.~6, 60008,
  \doi{10.1209/0295-5075/132/60008}.

\bibitem{Dudka23}
Dudka~M., Krasnytska~M., Ruiz-Lorenzo~J.~J., Holovatch~{\relax Yu}., J. Magn.
  Magn. Mater., 2023, \textbf{575}, 170718, \doi{10.1016/j.jmmm.2023.170718}.

\bibitem{Privman91}
Privman~V., Hohenberg~P.~C., Aharony~A., Phase Transitions and Critical
  Phenomena, Vol. 14, Domb C., Lebowitz~J.~L.~(Eds.), Academic Press, New York,
  1991.

\bibitem{Stanley72}
Hankey~A., Stanley~H.~E., Phys. Rev. B, 1972, \textbf{6}, 3515--3542,
  \doi{10.1103/PhysRevB.6.3515}.

\bibitem{Stanley99}
Stanley~H.~E., Rev. Mod. Phys., 1999, \textbf{71}, S358--S366,
  \doi{10.1103/RevModPhys.71.S358}.

\bibitem{Delfino98}
Delfino~G., Phys. Lett. B, 1998, \textbf{419}, No.~1, 291--295,
  \doi{10.1016/S0370-2693(97)01457-3}.

\bibitem{CASELLE1998613}
Caselle~M., Hasenbusch~M., Nucl. Phys. B Proc. Suppl., 1998, \textbf{63},
  No.~1, 613--615, \doi{10.1016/S0920-5632(97)00848-7}.

\bibitem{ENGELS2003277}
Engels~J., Fromme~L., Seniuch~M., Nucl. Phys. B, 2003, \textbf{655}, No.~3,
  277--299, \doi{10.1016/S0550-3213(03)00085-3}.

\bibitem{Kenna11}
Gordillo-Guerrero~A., Kenna~R., Ruiz-Lorenzo~J.~J., J. Stat. Mech., 2011,
  \textbf{2011}, No.~09, P09019, \doi{10.1088/1742-5468/2011/09/P09019}.

\bibitem{Davies83}
Davies~R.~A., Pepper~M., Kaveh~M., J. Phys. C, 1983, \textbf{16}, No.~10, L285,
  \doi{10.1088/0022-3719/16/10/006}.

\bibitem{STOOP91}
Stoop~R., Peinke~J., Parisi~J., Physica D, 1991, \textbf{50}, No.~3, 405--411,
  \doi{10.1016/0167-2789(91)90007-V}.

\bibitem{McLachlan98}
McLachlan~D.~S., Heiss~W.~D., Chiteme~C., Wu~J., Phys. Rev. B, 1998,
  \textbf{58}, 13558--13564, \doi{10.1103/PhysRevB.58.13558}.

\bibitem{Dulan19}
Walsh~D.~J., Guironnet~D., PNAS, 2019, \textbf{116}, No.~5, 1538--1542,
  \doi{10.1073/pnas.1817745116}.

\bibitem{Krummenacher22}
Krummenacher~M., Steinhauser~M.~O., J. Chem. Phys., 2022, \textbf{157}, No.~15,
  154904, \doi{10.1063/5.0108479}.

\bibitem{vonFerber11}
von Ferber~C., Folk~R., Holovatch~{\relax Yu}., Kenna~R., Palchykov~V., Phys.
  Rev. E, 2011, \textbf{83}, 061114, \doi{10.1103/PhysRevE.83.061114}.

\bibitem{Krasnytska16}
Krasnytska~M., Berche~B., Holovatch~{\relax Yu}., Kenna~R., J. Phys. A: Math.
  Theor., 2016, \textbf{49}, No.~13, 135001,
  \doi{10.1088/1751-8113/49/13/135001}.

\bibitem{Karsch23}
Karsch~F., Schmidt~C., Singh~S., Phys. Rev. D, 2024, \textbf{109}, 014508,
  \doi{10.1103/PhysRevD.109.014508}.

\bibitem{Kac52}
Berlin~T.~H., Kac~M., Phys. Rev., 1952, \textbf{86}, 821--835,
  \doi{10.1103/PhysRev.86.821}.

\bibitem{Leone02}
Leone~M., V{\'{a}}zquez~A., Vespignani~A., Zecchina~R., Eur. Phys. J. B, 2002,
  \textbf{28}, 191--197, \doi{10.1140/epjb/e2002-00220-0}.

\bibitem{Igloi02}
Igl\'oi~F., Turban~L., Phys. Rev. E, 2002, \textbf{66}, 036140,
  \doi{10.1103/PhysRevE.66.036140}.

\end{thebibliography}

\ukrainianpart

\title{Модель Ізінга зі змінною силою спіна на безмасштабній
	мережі: скейлінгові функції та співвідношення критичних амплітуд}
\author{М. Красницька\refaddr{label1,label2,label3}}
\addresses{
	\addr{label1} Інститут фізики конденсованих систем НАН України, Львів, 79011, Україна
	\addr{label2} Співпраця $\mathbb{L}^4$ і Коледж докторантів `Статистична фізика складних систем', Ляйпціґ-Лотарингія-Львів-Ковентрі, Європа
	\addr{label3} Haiqu, Inc., вул. Шевченка, 120 Г, 79039, Львів, Україна
}

\makeukrtitle

\begin{abstract}
	Нещодавно була розроблена нова модель для опису впорядкування в системах агентів, які, хоча збігаються у своїй бінарності (тобто, зберігають стандартні властивості моделі Ізінга --- значення ''+`` або ''--'', ''вгору'' або ''вниз'',
	``так'' чи ``ні''), але відрізняються силою агента [Krasnytska et al., J. Phys. Complex., 2020, \textbf{1}, 035008]. Ми дослідили Модель для конкретного випадку, а саме, коли агенти розташовані на вузлах безмасштабної мережі і сила агента є випадковою
	змінною, яка регулюється степенево спадним законом. Для відпаленої мережі точний розв'язок демонструє багату фазову діаграму з різними типами критичної поведінки та новими класами універсальності. Ця стаття продовжує вищезазначені дослідження та описує поведінку скейлінгових функцій та універсальних критичних співвідношень амплітуд для Моделі на безмасштабній мережі.

 \keywords фазові переходи, модель Ізінга, універсальність, скейлінгові функції, співвідношення критичних амплітуд, складна мережа
 
\end{abstract}

\lastpage
\end{document}